\documentclass[aps,amsfonts,amsmath,prd,preprint,tightenlines,nofootinbib]{revtex4}

\def\be{\begin{equation}}
\def\ee{\end{equation}}

\begin{document}

\title{Are there Boltzmann brains in the vacuum?}
\author{Matthew Davenport}
\email{Matthew.Davenport@tufts.edu}
\author{Ken D. Olum}
\email{kdo@cosmos.phy.tufts.edu}
\affiliation{Institute of Cosmology, Department of Physics and
Astronomy, Tufts University, Medford, MA 02155}

\begin{abstract}

``Boltzmann brains'' are human brains that arise as thermal or quantum
fluctuations and last at least long enough to think a few thoughts.
In many scenarios involving universes of infinite size or duration, Boltzmann
brains are infinitely more common than human beings who arise in the
ordinary way.  Thus we should expect to be Boltzmann brains, in
contradiction to observation.  We discuss here the question of whether
Boltzmann brains can arise as quantum fluctuations in the vacuum.
Such Boltzmann brains pose an even worse problem than those arising as
fluctuations in the thermal state of an exponentially expanding
universe.  We give several arguments for and against inclusion of
vacuum Boltzmann brains in the anthropic reference class, but find
neither choice entirely satisfactory.

\end{abstract}

\maketitle

\section{Introduction}

The basic idea of anthropic reasoning is that we should expect
ourselves to be typical among some set of intelligent beings, called
the reference class.  This is codified, for example, in Vilenkin's
\cite{gr-qc/9406010} ``principle of mediocrity'' or Bostrom's
\cite{Bostrom:book} ``self-sampling assumption''.  As an example of
this, we can take someone who just bought a lottery ticket.  He should
consider himself in equivalent circumstances to everyone else who has
bought a ticket, assuming he has no inside information on the outcome
of the draw.  There are then two subgroups that he can be in, people
with winning tickets and those with losing tickets, but there are a
million members in the latter class and only one in the former.  Thus,
the person can assume that he is much more likely a member of the
larger subgroup through anthropic reasoning.  

While the precise specification of the reference class is a matter of
considerable uncertainty, it seems clear that we should include all people
subjectively indistinguishable from us \cite{Bostrom:book}.  After
all, by assumption we have no way to know that we are not those
people.  For the purpose of this paper, this minimal reference class
is sufficient.

To make things simple, let us assume that the universe is very large
but finite in spatial extent.  Let us suppose that the observed dark
energy is in fact a cosmological constant. The accelerating expansion
will then dilute away all the matter and radiation that fill the
universe today, the universe will approach empty de Sitter space, and
the state of that space will approach the Bunch-Davies
\cite{Bunch:1978yq} vacuum.  In that state, it is possible for objects
to nucleate.  For example, an electron-positron pair can form
spontaneously, separated by more than the horizon distance
\cite{Gibbons:1977mu}.  Similarly, it is also possible for a group of
these particles to form together with their respective antiparticles
so that the conservation laws are obeyed.  With enough particles
forming together in various configurations, it is possible to produce
any physical object.  In particular, it is possible for there to
spontaneously appear a brain that is in exactly the state of your
brain at this moment, and thus is apparently indistinguishable from
you.  The chance per unit space-time volume for such an object to
appear is infinitesimal, but it is nonzero.  Since the de Sitter phase
will exist forever, the number of such ``Boltzmann brains''
\cite{Barrow:book,Albrecht:2004ke}, will grow without bound, while the
number of normal observers is finite.  Thus by anthropic reasoning you
should believe with probability 1 that you are one of the Boltzmann
brains.

Of course, no one really believes that he is a Boltzmann brain.  For
one thing, to believe so would bring up the problem of skepticism in a
severe form.  Believing you were a Boltzmann brain would give you a
strong reason to deny the existence of the external world.  Hume's
answer to skepticism was that even though we cannot affirm the
existence of the external world, in order to live from day to day we
must act as though it really exists.  But here, assuming the existence
of a particular external world has led us to conclude that in fact the
external world does not exist (at least in the form we appear to
observe it).  This is unacceptable because it prevents us from
affirming the existence of the external world as Hume says we must.
Thus by Hume's analysis, we must reject any theory according to which
we should be Boltzmann brains.

Furthermore, there is a simple test to see whether you are a Boltzmann
brain.  Wait 1 second and see if you still exist.  Most Boltzmann
brains are momentary fluctuations.  So the prediction of the above
argument is that you will vanish in the next second.  When you don't,
you conclude that this argument has made a severely wrong
prediction.\footnote{If you are concerned with the fact that you could
  never observe your own ceasing to exist, you can change the argument
  to say that the theory that you are a Boltzmann brain predicts that
  your observations of the external world are coherent only by chance,
  and that subsequent observations will not remain coherent.}

What could be wrong?  One possibility is that our assumption of a
finite universe is not correct.  But if the universe is infinitely
large but homogeneous, and otherwise as specified above, the argument
seems to be the same.  In this case, there are infinitely many
ordinary observers and infinitely many Boltzmann brains.  But there
seems to be a simple solution of computing the number of different
kinds of observers who ever exist in a given comoving volume.  As long
as one takes a sufficiently large volume and does not choose it in a
special way based on the existence of observers, the resulting density
of observers will not depend on the choice of volume.  Again this
density is finite for ordinary observers but infinite for Boltzmann
Brains.

Another possibility is that our universe will be short-lived, so that
there will not be a long period of time during which Boltzmann brains
could form.  Using this argument, Page
\cite{hep-th/0510003,hep-th/0610079} set several bounds on the future
age of the universe, the most striking being that the universe would
last less than 20 billion years.

To avoid such unfortunate conclusions, we can take advantage of the
fact that compact objects such as Boltzmann brains aren't the only
things that can nucleate in de Sitter space.  There can also nucleate
horizon-size inflating regions \cite{Garriga:1997ef}, which then go on
to thermalize and produce ordinary observers (and also large new
regions of de Sitter space in which more Boltzmann brains can form.)
More generally, our universe may be part of a complex multiverse
(formed, for example, in a string theory landscape
\cite{Susskind:2003kw}), in which new and different universes nucleate
in inflating regions.  To compute probabilities in any such case
requires choosing a measure on the infinite and inhomogeneous
spacetime.  There is no agreement on the appropriate measure
\cite{Aguirre:2006ak}, but it seems possible that the correct measure
will not give an unreasonable probability for us to be Boltzmann
brains\footnote{In fact, this consideration is sometimes used as
  evidence of which measure is correct.}, so perhaps the solution to
the paradox lies there.

But in addition to those that nucleate in de Sitter space, there is
another kind of Boltzmann brain.  Suppose that the dark energy is a
dynamical field (such as quintessence) which will eventually decay
away.  In that case, the universe will become increasingly empty and
will have at the end an infinite period of what is essentially
Minkowski space in the vacuum state.  In Minkowski space, nucleation
is not possible, because of conservation of energy.  However,
it may still be possible for observers indistinguishable from us
to arise as vacuum fluctuations in this infinite volume of
Minkowski spacetime.  Such observers will be the subject of this paper.

In one sense, the vacuum is just the vacuum, a pure state, and it
contains no observers.  But what happens if we consider not the entire
spacetime but just the state of some particular region $R$ at a given
time.  The situation is analogous to a system of two coupled harmonic
oscillators in the ground state.  The ground-state wavefunction of that
system can be expressed in terms of eigenstates of the harmonic
oscillators separately, as
\be
|0\rangle = \sum_n a_n |n\rangle_1 \otimes |n\rangle_2
\ee
where $|n\rangle_i$ denotes the $n$th excited state of oscillator $i$, and
$a_n$ is a numerical coefficient.  Similarly, one can write the vacuum
of Minkowski space as a superposition
\be
|0\rangle = \sum_\alpha a_\alpha |\alpha\rangle_R \otimes |\alpha\rangle_{\bar R}
\ee
The sum is over all possible states $|\alpha\rangle_R$ that could
possibly be found inside $R$, with corresponding states found outside
$R$.  The factor $a_\alpha$ is some nonzero coefficient for each
state.

This sum includes all states.  In particular it includes states in
which $R$ contains a Boltzmann brain with all your present thoughts.
If one interprets $|a_\alpha|^2$ as the probability that the region
$R$ contains an observer indistinguishable from you, one gets a
different version of the Boltzmann brain paradox.  Since the spacetime
volume is infinite, there are infinitely many regions equivalent to
$R$.  As long as $a_\alpha$ corresponding to a brain is nonzero,
there are infinitely many Boltzmann brain versions of you.  In
contrast to the de Sitter case, in the infinite Minkowski region one
cannot have new universes appear, so that way out of the paradox is
closed.

If one takes this problem seriously, perhaps one could conclude that
the dark energy will never decay.  But even then, this second type of
Boltzmann brain still exists.  The low-temperature Bunch-Davies vacuum
of de Sitter space is locally very similar to the zero-temperature
vacuum of Minkowski space, and the amplitude to have a Boltzmann brain
in each space-time volume is essentially the same.

We can compare the probability (i.e., the $|a_\alpha|^2$) of a
Boltzmann brain in the Minkowski space vacuum wave function with the
chance of a Boltzmann brain nucleating in de Sitter space.  The
probability to nucleate a compact object\footnote{Here we neglect the
  entropy of the object, i.e., the fact that there are many possible
  brains with different microphysical states, even though they are
  indistinguishable at the level of thoughts.  The nucleation rate
  should be multiplied by the number of such possibilities, but the
  correction to the huge exponent would be relatively tiny.} in de
Sitter space is given by \cite{Basu:1991ig} \def\pds{p_{\text{dS}}}
\be
\pds \propto \exp\left(-\frac{2 \pi Mc^2}{\hbar H}\right)\,,
\ee
where $H = \sqrt{(8 \pi/3) G \rho_\Lambda}$ is the Hubble constant of
de Sitter space, and $c$ is the speed of light.  Using an approximate
value $M=1$ kg, and the present dark energy density $\rho_\Lambda \sim
10^{-29} \text{g}/\text{cm}^3$, we get
\be
\pds \sim \exp\left(-10^{69}\right)\,.
\ee
Considering the uncertainty about the exponent, there is no point in
calculating the prefactor.

In contrast, Page \cite{hep-th/0510003} computes the probability of a
brain in a specific sequence of states lasting for time $t$ in a given
region in the Minkowski vacuum state to be proportional to $\exp(-M t
c^2/\hbar)$, where $M$ is the mass of the brain.  Again using $M =
1$kg, and $t= 0.1 s$, the ``time to have a thought'', he finds
probability \def\pvac{p_{\text{vac}}}
\be
\pvac \sim \exp(-10^{50})\,.
\ee

Thus $\pvac$ is enormously larger than $\pds$.  The ratio is
$\pvac/\pds = \exp(10^{69}-10^{50}) \approx \exp(10^{69})$,
i.e., $\pvac/\pds$ is essentially indistinguishable from $1/\pds$.  An
explanation of the Boltzmann brain paradox in terms of the measure on
the multiverse will have a harder time solving the problem if a
Boltzmann brain production rate proportional to $\pvac$ must be
overcome than if we need overcome only an infinitesimally smaller rate
depending on $\pds$.

Thus it seems important to determine which kinds of Boltzmann brains
should be considered as part of the reference class.  One might try to
argue that no Boltzmann brain should be considered, for example
because in either case the objects are sectors of a pure-state wave
function.  But this does not seem feasible.  Assuming the usual
scenario of inflation is correct, our galaxy condensed from an early
quantum fluctuation in the field driving inflation.  Thus we ourselves
are a sector of a pure-state wave function describing the inflating
universe.  Since we put ourselves in the reference class, it appears
that arising from a quantum fluctuation is not a disqualification.  If
eternal inflation is correct, our whole universe was nucleated as a
bubble in de Sitter space, and there is an even closer analogy to the
nucleation of Boltzmann brains.

Vacuum fluctuations, however, are somewhat different.  While
nucleating objects can last forever, vacuum fluctuations in Minkowski
space have a finite lifetime and can never decohere, since the vacuum
wave function must remain pure.  We shall discuss these issues below.

\section{Deluded Boltzmann brains}

In the simplest (and thus vastly most common) case, the Boltzmann
brain is not part of a ``Boltzmann universe'' (the original idea of
Boltzmann \cite{Boltzmann}) but rather exists on its own with a
minimum amount of life support necessary to think a few thoughts.
Thus if we are Boltzmann brains, all our past experiences are just an
elaborate illusion with no connection to the external world.  Thus our
observations of the universe and our knowledge of physical laws are
just delusions, and we have no real knowledge of anything outside
ourselves.  But it is just such knowledge of the supposed external
world which has led us to formulate the Boltzmann brain paradox in
the first place.  If we abandon it, we will have no understanding of
the real world and thus no possibility to understand the likelihood
that we are Boltzmann brains.

So let us state the problem more carefully.  Suppose, based on our
observations, we come up with some theory, say ``theory X'' of physics
and cosmology.  We want to know whether theory X is correct, so we
compare our observations with what it predicts.  If theory X has a
Boltzmann brain problem, then it predicts (usually with probability 1)
that we are Boltzmann brains.  The vast majority of Boltzmann brains
do not have correct observations of the external world, so their
observations would not be in accord with theory X, as ours are.
Furthermore the great majority of Boltzmann brains would vanish
immediately, as discussed above.  Thus theory X leads to
high-confidence predictions at odds with observation.

We do not in this analysis consider Boltzmann brains arising under
some other theory Y but falsely believing they exist under theory X.
Of course such Boltzmann brains could exist, but we have no way of
meaningfully discussing them, because we have no way of knowing all
possible theories Y that might have Boltzmann brains.

\section{Arguments for and against vacuum Boltzmann brains}

\subsection{Turing test}

Minkowski-space vacuum fluctuations are local, in the sense that,
while consideration of the state restricted to a local region gives a
nonzero amplitude for a Boltzmann brain, the state of the entire space
is just the unchanging, pure vacuum.  Thus it is not possible for
a brain arising as a vacuum fluctuation to send out any durable signal
of its existence.  If it could, the global state would now consist (at
least) of the vacuum plus the outgoing signal.

(Vacuum fluctuations in de Sitter space are very similar, except if
the fluctuation or its effects extend across the de Sitter horizon.
In that case, we have not the vacuum-fluctuation case that is
analogous to Minkowski but the different case of nucleation, as
discussed above.)

Thus a vacuum-fluctuation Boltzmann brain can never pass the ``Turing test''.
According to Turing \cite{Turing}, to determine whether something has
human intelligence we should allow a human interrogator to ask it
questions and see if its answers are indistinguishable from those of a
real human.  But the vacuum Boltzmann brain can never communicate with
any external entity, so the test cannot be done \cite{Gott:2008ii}.

More generally, the thoughts of a vacuum-fluctuation brain can never
have any impact on anything.  Thus if one believes that ``a difference
that makes no difference is no difference'' (William James) then the
existence of a vacuum Boltzmann brain can never matter for any
purpose.  These arguments appear to motivate the exclusion of
vacuum-fluctuation brains from the reference class.

On the other hand, the same argument would apply to an ordinary
observer enclosed in a perfectly sealed box (which we imagine here to
be sealed even against gravitational waves).  His thoughts could never
affect anything outside the box.  If we were sealed in such a box,
especially without our knowledge, it seems very strange to declare that
we should no longer be members of the reference class.

In fact, if we deny the reality of the vacuum-fluctuation brain, it
appears that we are denying the famous argument of Descartes.  A brain
could be thinking ``cogito ergo sum,'' at least in the sense that the
vacuum wave function is made up in part of a brain thinking that
thought.  But when we deny its reality, we say that this ``Boltzmann
Descartes'' is wrong, at least to the extent that existence means
membership in the reference class.  This seems a powerful argument in
favor of including vacuum-fluctuation brains in anthropic reasoning.

\subsection{Decoherence}

It is often argued that quantum mechanical decoherence produces the
distinction between quantum and classical behavior.  Decoherence means
irreversible coupling to the environment.  For example, if the
electron in a two-slit experiment is permitted to interact with a bath
of background photons, there will be no interference pattern.  The
quantum state of the electron necessary to produce interference is
entangled with the photons and transported away from the apparatus.
In principle one could track down these photons and recover the
coherence and thus the interference pattern, but in practice that
would not be possible.  Decoherence explains why we never see people's
brains in superposition states, or cats in a superposition of live and
dead.

Since a Minkowski vacuum fluctuation can never send out any durable
signal, it can never be decoherent.  This is different from the
situation in de Sitter space.  A nucleating brain in de Sitter space
can send out signals which travel outward forever, and can thus
decohere in the usual way.  Even though the entire de Sitter space
might be in a pure quantum state, the existence of horizons allows for
decoherence.

One could perhaps say that events that are not decoherent never
actually take place \cite{arXiv:0812.0605}, and thus that the
``thoughts'' of vacuum Boltzmann brain are not real thoughts.  Or one
could say that non-decoherent observers should never be counted in
anthropic reasoning.  This is similar to the idea of Gott
\cite{Gott:2008ii} that Boltzmann brains should not be counted if they
are observer-dependent, i.e., if one observer (here, one who
looks only inside a given region) finds them to exist, while another
(here, one who observes the entire Minkowski vacuum) does not.

Such strategies, however, have some unpalatable consequences.  If we
introduce into the vacuum a physical measuring apparatus that actually
performs a measurement of the content of a certain region, it could
potentially find a vacuum Boltzmann brain there.  In this case, the
Boltzmann brain would acquire decoherence and permanent existence
through the process of measurement.  (The measurement apparatus would
have to be able to supply a sufficient amount of energy to allow the
Boltzmann brain to exist without violating energy conservation.)  The
above arguments against the reality of this brain would no longer
operate.  However, the brain could have thought many thoughts before
the time of measurement.  Thus to go this route requires accepting the
idea that the reality of the thoughts is not determined at the time of
the thinking, but only later at the time of (possible) decoherence.
On the other hand, perhaps this idea is no worse than the generally
accepted idea that the reality of the spin of an electron pointing in
a certain direction can be established only later when the spin is
measured.

\subsection{Reversibility}

To gain some insight into the question of vacuum Boltzmann brains, we
can imagine a computer that could exist as a vacuum fluctuation.  It
cannot send out any signal or interact with the external world.  By
analogy with a vacuum state, the quantum-mechanical states of the
computer before and after its operation should be the same.  Thus the
computer cannot store any of the results of its operation.

These criteria correspond exactly to the operation of a reversible
computer that is reversed.  It starts in some initial state, performs
some computation to end in a ``final'' state, and is then reversed to
go backward from the final to the initial state, all while maintaining
quantum coherence.

Such a computer could be put, for example, in one of the paths in a
double slit experiment.  An electron is incident on a plate with two
slits.  Through the lower slit, the electron travels unimpeded to a
screen.  Through the upper slit, the electron triggers a complex
reversible computation.  The reversible computer arrives at the final
state and then returns to the initial state, eventually reproducing
the electron with the same quantum state as it had originally.  The
electron then travels onto the same screen, where we see an
interference pattern.\footnote{We can put a delay in the lower path so
  that the electron arrives at the screen at the same time regardless
  of which way it goes.}  The existence of the interference pattern
guarantees that the computation is not decoherent.

One may now ask whether the computation was performed, was not
performed, or was performed with 50\% probability.  Under normal
circumstances, this question would be meaningless.  Asking ``was the
computation performed?'' is analogous to asking ``which path did the
electron take?''.  It is not a meaningful question.  

However, anthropic reasoning appears to have a ``super power'' to make
the answers to otherwise meaningless questions matter.  For example,
suppose that we create 1000 boxes involving double-slit experiments
and reversible computers, where the computation of the computers
simulates the thinking of the human brain.  Suppose we create, in
addition, 10 ordinary computers that simulate human thought.  Suppose
you know that you are one of these 1010 computers.\footnote{To sharpen
  the question, one could imagine that when you go to sleep tonight
  the entire state of your brain will be copied into each of these
  computers, and your organic body destroyed.  Thus when you wake up,
  you will be one of the computers, but you don't know which one.}
Now the question of whether the computation is really performed
becomes relevant.  If you think it is performed in every box, then you
should think the odds of being in one of the double-slit experiments
are 100:1.  If you think it is performed in half the boxes, you should
think 50:1, while if you think that it is never performed, or that a
reversing computer never counts in anthropic calculations, you should
think it certain that you are one of the ordinary computers.

One could perform a slightly different version of this experiment by
stopping some of the computers at the reversal point and opening the
boxes to permit decoherence.  Perhaps these computers should be
included in the reference class, even though the actually reversed
ones should not.  As discussed in the previous section, this idea has
unpalatable consequences.

The reversal of the computers in this example has the strange property
that the computers must ``unthink'' their thoughts to arrive back at
the initial state.  Such ``unthinking'' consists of stepping backward
through the same sequence of states that represents thinking.  If one
considers the first half of the process as an observer in the
reference class, one could consider the second half of the process as
a time-reversed observer in the reference class.  Thus it appears that
vacuum Boltzmann brains have a 50\% chance to be mistaken about the
direction of time in the universe of which they are part.  This
follows, essentially, from the time-reversibility of the vacuum state.

\section{Conclusion}

Boltzmann brains pose a problem for many theories of physics and
cosmology.  If Boltzmann brains found as vacuum fluctuations in
Minkowski space (or fluctuations whose effects exist only within a de
Sitter space horizon) are included in the reference class, the
``Boltzmann brain problem'' is even more severe.  However, such
Boltzmann brains have anomalous features: they can never be
decoherent, they must be reversible, and they can never communicate
with the outside world and thus cannot pass the Turing test.  In fact,
to a global observer, these brains do not exist at all: there is only
the vacuum wave function.  These features motivate the exclusion of
Boltzmann brains from the reference class.

However, excluding vacuum-fluctuation Boltzmann brains from the
reference class has its own unfortunate features.  It appears that we
cannot decide whether an observer is to be counted until we know
whether he decoheres, which might not be until much later.
Furthermore, one may argue that such exclusion violates the Cogito: we
deny the existence of a brain that is thinking.

In the end, we can conclude only that a completely satisfactory
explanation of the proper handling of Boltzmann brains arising as
vacuum fluctuations is yet to be constructed.

\section*{Acknowledgments}

We thank Michael Salem, Ben Shlaer, and Alex Vilenkin for helpful
discussions.  This research was supported by in part by grant
MGA-08-006 from The Foundational Questions Institute (fqxi.org).

\bibliography{no-slac,vacuum}

\end{document}